\documentclass[conference, 9pt]{IEEEtran}
\IEEEoverridecommandlockouts
\usepackage{amsmath,graphicx}
\usepackage{rotating}
\usepackage{epstopdf}
\usepackage{cite}
\usepackage{amsmath, amssymb, amsthm}
\usepackage{breqn}
\usepackage{graphicx}
\usepackage{stfloats}
\usepackage{url}
\usepackage{xcolor}
\usepackage{multirow}
\usepackage{bm}
\usepackage{microtype}
\usepackage{fixltx2e}
\usepackage{longtable,eqnarray}
\usepackage{mathtools,soul}
\usepackage{adjustbox}
\usepackage{footnote}
\usepackage{algorithm}
\usepackage{subfig}
\usepackage{algpseudocode}
\def\BibTeX{{\rm B\kern-.05em{\sc i\kern-.025em b}\kern-.08em
    T\kern-.1667em\lower.7ex\hbox{E}\kern-.125emX}}

\makeatletter
\def\ps@IEEEtitlepagestyle{%
  \def\@oddfoot{\mycopyrightnotice}%
  \def\@oddhead{\hbox{}\@IEEEheaderstyle\leftmark\hfil\thepage}\relax
  \def\@evenhead{\@IEEEheaderstyle\thepage\hfil\leftmark\hbox{}}\relax
  \def\@evenfoot{}%
}
\def\mycopyrightnotice{%
  \begin{minipage}{\textwidth}
  \centering \scriptsize
  Copyright~\copyright~2025 IEEE. Personal use of this material is permitted. Permission from IEEE must be obtained for all other uses, in any current or future media, including reprinting/republishing this material for advertising or promotional purposes, creating new collective works, for resale or redistribution to servers or lists, or reuse of any copyrighted component of this work in other works.
  \end{minipage}
}
\makeatother

\begin{document}
\title{Robust Fixed-Filter Sound Zone Control with Audio-Based Position Tracking
{\footnotesize \thanks{\textsuperscript{*}Equal contribution by Sankha Subhra Bhattacharjee and Andreas Jonas Fuglsig. This work is supported by Huawei Technologies Co., Ltd.}}
}
\author{\IEEEauthorblockN{Sankha~Subhra~Bhattacharjee\textsuperscript{*}, Andreas~Jonas~Fuglsig\textsuperscript{*}, Flemming~Christensen,\\ Jesper~Rindom~Jensen and Mads~Græsbøll~Christensen} \IEEEauthorblockA{\textit{Department of Electronic Systems,}\\ \textit{Aalborg University, 9220 Aalborg, Denmark}\\ Email: \{ssbh,ajf,fc,jrj,mgc\}@es.aau.dk}}

%

%
\maketitle

\begin{abstract}
Performance of sound zone control (SZC) systems deployed in practical scenarios are highly sensitive to the location of the listener(s) and can degrade significantly when listener(s) are moving. This paper presents a robust SZC system that adapts to dynamic changes such as moving listeners and varying zone locations using a dictionary-based approach. The proposed system continuously monitors the environment and updates the fixed control filters by tracking the listener position using audio signals only. 
To test the effectiveness of the proposed SZC method, simulation studies are carried out using practically measured impulse responses. These studies show that SZC, when incorporated with the proposed audio-only position tracking scheme, achieves optimal performance when all listener positions are available in the dictionary. Moreover, even when not all listener positions are included in the dictionary, the method still provides good performance improvement compared to a traditional fixed filter SZC scheme. 


\end{abstract}


\section{Introduction}

The goal of Sound Zone Control (SZC) is to create personalized audio environments for different listeners in a shared space, with minimal interference between regions~\cite{betlehem2015personal}. SZC has found application in various fields, including smartphones \cite{jeon2015active}, outdoor concerts \cite{heuchel2020large},
car audio systems\cite{vindrola2021use}, and home entertainment systems \cite{galvez2014personal}.

The basic principle of SZC involves designing control filters for multiple loudspeakers to generate a sound field such that so-called Bright Zone(s) (BZ/BZs) receive the intended audio signals, while the audio content is attenuated in so-called Dark Zone(s) (DZ/DZs). Some of the most common SZC methods include Acoustic Contrast Control (ACC), where the ratio of acoustic energies between the BZ and DZ are maximized \cite{jones2008personal, cai2013design}; Pressure Matching (PM), which aims to minimize both signal distortion in the BZ and energy in the DZ \cite{poletti2008investigation, olivieri2017generation,chang_sound_2012}; and Variable-Span Trade-off (VAST), which provides a balance between the ACC and PM solutions \cite{lee2018unified, nielsen2018sound, shi2021generation, brunnstrom2022variable}. Traditional SZC methods rely on fixed filters designed based on the acoustic properties of the environment and the positions of the control points. However, changes in the environment, such as temperature variations\cite{olsen_sound_2017, coleman_acoustic_2014}, loudspeaker and microphone movements~\cite{coleman_acoustic_2014,park_acoustic_2013}, alterations in zone sizes\cite{jacobsen_living_2023} and moving listeners or zones \cite{moller_moving_2020} can degrade the performance of SZC systems if these do not adapt. The dynamic factors introduce challenges in maintaining the desired sound field, as the pre-designed filters may no longer be optimal under the new conditions. In this work, we particularly focus on the challenge of moving listeners or zones.

Various approaches have addressed some of these challenges with changing environments. For instance, regularization techniques\cite{coleman_acoustic_2014, elliott_robustness_2012}
have been employed to enhance the robustness of SZC systems against environmental changes. However, these do not take into account any new information about the changes.
In addition, adaptive filters have been proposed in combination with virtual microphones\cite{vindrola2021use} to track changes in the environment. Similarly, a moving horizon framework \cite{moller_moving_2020} has been used to predict and adapt to changes. However, these have mainly been studied for frequencies below $1$\,kHz. 
As an alternative to directly adapting the filters, dictionary-based approaches have been popular in the field of active noise control, to handle moving listeners\cite{han_combination_2019, jung_combining_2017, elliott_head_2018, chang_multi-functional_2022, xiao_ultra-broadband_2020, buck_performance_2018,shi_selective_2022}. These methods use external sensors, such as cameras or infrared sensors, to detect the listener's position and select the appropriate filter from a pre-computed dictionary. However, such additional sensors may not always be available or feasible to install in a pre-existing system.

In this paper, we propose an audio-only method for tracking the listener position and subsequent selection of the best filter solution from a dictionary of pre-trained fixed filters. The proposed method uses a set of observation microphones (different from the BZ and DZ control points) to monitor the sound field, along with a dictionary of pre-measured Impulse Responses (IRs) from the SZC loudspeakers to the set of observation microphones for different listener positions on a grid to select the best filter.
During frame-wise SZC deployment, the signals recorded at the observation microphones are matched with internally computed estimates of the observation signals at the different positions in the dictionary of pre-measured IRs, where signals are matched for similarity using the normalized cosine similarity (NCS) metric \cite{luo2018cosine,tan_introduction_2014}. 
The dictionary listener position that gives the maximum value of the NCS is considered as an estimate of the true listener position and the corresponding control filter solution is selected from the control filter dictionary for SZC. The performance of the proposed method is compared for two different cases, one where all possible listener positions lie on the measurement grid points in the dictionary and one where listener positions can lie outside the measurement grid points.

\section{Sound Zone Control Background}\label{sec:sond_zone_control}

\begin{figure}[]
  \centering
  \includegraphics[trim={0 3mm 0 0}, clip,width=\linewidth]{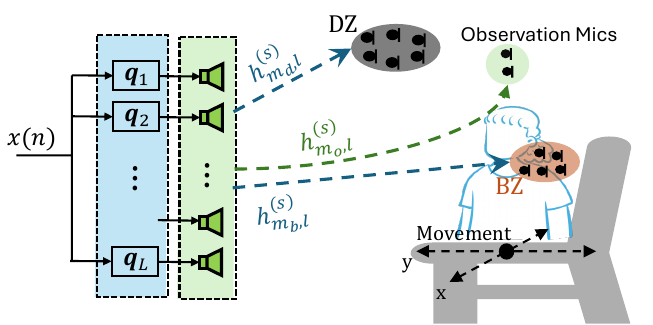}
  \caption{SZC scenario with a BZ next to the ear of a person moving on a chair and a DZ off to the side, with microphones placed in the room for monitoring/observing the listener position.}
  \label{fig:SZC_scenario}
\end{figure}

As an example, we consider a SZC problem, cf. Fig.~\ref{fig:SZC_scenario}, where a listener moves on a chair, 
limited to translation motions.
The goal is to create and maintain a BZ with $M_B$ Control Points (CPs) next to the listener's ear, i.e., the BZ moves along with the listener, while minimizing sound energy in the DZ with $M_D$ CPs. 
The SZC system also has $M_O$ observation
microphones, different from the BZ and DZ CPs, for tracking the listener's position during SZC deployment and uses $L$ loudspeakers for creating the zones.
For a given listener position, $s$, we let ${\bm{h}^{(s)}_{m_b,l} \in \mathbb{R}^{K}}$ represent the IR from the $l^{\text{th}}$ loudspeaker to the $m_b^{\text{th}}$ microphone in the BZ. Similarly, ${\bm{h}^{(s)}_{m_d,l} \in \mathbb{R}^{K}}$ and ${\bm{h}^{(s)}_{m_o,l} \in \mathbb{R}^{K}}$ represent the IR from the $l^{\text{th}}$ loudspeaker to the $m_d^{\text{th}}$ CP in the DZ and the $m_o^{\text{th}}$ 
observation
microphone, respectively. The sound is controlled by the control filters $\bm{q}^{(s)}_{l} \in \mathbb{R}^{J}$, optimal for listener position $s$, applied to the loudspeakers.

\subsection{Fixed Filter Sound Zone Control}
When deriving SZC filters, we commonly assume the input signal is $x[n]=\delta[n]$, i.e., a deterministic spectrally white signal \cite{galvez2015time, moles2020personal, moller2016sound, bhattacharjee2023study}. The optimal filters are thus time-invariant and only optimal for a particular position, $s$. Following this assumption, the sound reproduced at the $m^{\text{th}}$ microphone in either the BZ, DZ or at the 
observation
microphones can be expressed as the vector
${\bm{p}_m^{(s)}= \sum\nolimits_{l=1}^{L} \bm{h}^{(s)}_{m,l} * \bm{q}^{(s)}_l=
  \sum\nolimits_{l=1}^{L} \bm{H}^{(s)}_{m,l}\bm{q}^{(s)}_l}$,
where $*$ denotes discrete time convolution, and $\bm{H}^{(s)}_{m,l}$ is the convolution matrix for the $m^{\text{th}}$ CP and $l^{\text{th}}$ loudspeaker IR $\bm{h}^{(s)}_{m,l}$ \cite{galvez2015time}.
Now, the sound signals reproduced at all BZ microphones are given by
\begin{flalign}
  \bm{p}^{(s)}_B = \left[\bm{p}^{(s)}_1, \ldots, \bm{p}^{(s)}_{M_B}\right]^{\textrm{T}} &= \bm{H}^{(s)}_B \bm{q}^{(s)} \in \mathbb{R}^{M_B(K+J-1)}
\end{flalign}
where 
${\bm{H}_B^{(s)}=\left\{\bm{H}^{(s)}_{m,l}\right\}_{m=1,\ldots,M_B,l=1,\ldots,L}}$ is a block matrix
and 
$ {\bm{q}^{(s)}= \left[\left(\bm{q}^{(s)}_1\right)^{\textrm{T}}, \ldots, \left(\bm{q}^{(s)}_{L}\right)^{\textrm{T}}\right]^{\textrm{T}}} \in \mathbb{R}^{LJ}$\cite{galvez2015time}.
Letting ${\bm{R}_B^{(s)} = \bm{H}^{(s)}_B\left(\bm{H}^{(s)}_B\right)^{\textrm{T}}}$ and $\bm{R}_D^{(s)} = \bm{H}^{(s)}_D\left(\bm{H}^{(s)}_D\right)^{\textrm{T}}$ be the covariance matrices for the BZ and DZ, respectively, we can derive a set of optimal control filters for any listener position, $s$, with known IRs.

\subsubsection{Acoustic Contrast Control (ACC)}
The optimal fixed filters for the ACC method can be derived by solving an optimization problem that maximizes the acoustic contrast between the BZ and DZ\cite{lee2018unified}.
%
The optimal solution to this problem, $\bm{q}_{\textrm{ACC}}^{(s)}$ is the 
principal eigenvector
of the Generalized Eigenvalue Decomposition (GEVD) problem ${\bm{R}_B^{(s)} \bm{q} = \mu' \left(\bm{R}_D^{(s)}+\lambda\bm{I}\right)\bm{q}}$ \cite{nielsen2018sound}, where $\lambda$ is a regularization parameter, $\bm{I}$ is the identity matrix, and $\mu' = \mu M_B/M_D$.


\subsubsection{Pressure Matching (PM)}
The PM method~\cite{poletti2008investigation} 
minimizes
the difference between the reproduced and a desired sound pressure in the BZ and DZ. We denote the desired sound pressure in the BZ and DZ, by $\bm{d}_B^{(s)}$ and $\bm{d}_D^{(s)}$, respectively, where $\bm{d}_B^{(s)}$ could be the sound pressure reproduced by a virtual source or a loudspeaker from the setup. The desired sound pressure in the DZ, $\bm{d}_D^{(s)}$, is typically set to the zero vector, $\bm{0}$, to minimize leakage into the DZ \cite{galvez2015time}.
The optimal fixed filters for the weighted PM method is derived by solving the following optimization problem
\cite{chang_sound_2012}
\begin{equation}
 \bm{q}_{\textrm{PM}}^{(s)}=
\arg\min_{\bm{q}\in \mathbb{R}^{LJ}} ~~ (1- \zeta)\lVert \bm{p}^{(s)}_B - \bm{d}_B^{(s)} \rVert_2^2 + \zeta\lVert \bm{p}^{(s)}_D \rVert_2^2 + \lambda\lVert\bm{q}\rVert_2^2,
\end{equation}
where $\lVert \cdot \rVert_2$ denotes $\ell_2$-norm, and $\zeta$ is a weighting factor that trades off fitting to the desired signal in BZ and minimizing energy in the DZ. The solution to this optimization problem is\cite{lee2018unified,galvez2015time}
\begin{equation}
  \bm{q}_{\textrm{PM}}^{(s)} = \left((1-\zeta)\bm{R}_B^{(s)}+\zeta\bm{R}_D^{(s)}+\lambda\bm{I}\right)^{-1}(1-\zeta)\bm{R}_B^{(s)}\bm{d}_B^{(s)}.
\end{equation}
Next, we use the ACC and PM methods to derive a dictionary of control filters, each optimal for a specific listener position, and propose an audio-only position tracking and control filter selection method.

\section{Dictionary based Sound Zone Control}\label{sec:dictionary_sound_zone_control}
We propose a robust SZC system that handles moving listeners without access to their position and without using microphones in the BZ or DZ at run-time (SZC deployed). We use a dictionary-based approach, where we pre-compute a dictionary, $\mathcal{D}$, of fixed filters for different listener positions. Then, at run-time, the observation microphones' signals are used to monitor the sound and select the best filter from the dictionary, based on similarity between the observed and estimated sound for each listener position in a dictionary $\mathcal{S}$.

%
%
We deploy the SZC using a time-domain (TD) frame-based approach with any general input signal,
where we divide it into non-overlapping frames of length $N$ as ${\bm{x}[\tau] = \left[x[\tau N - N + 1], \cdots, x[\tau N]\right] \in \mathbb{R}^{N}}$, where $\tau$ is the TD frame index. Each frame is then filtered by the SZC filters before playback. 
We use overlap-add and buffers to avoid frame boundary errors\cite{christensen_introduction_2019}, and denote the $l^{\text{th}}$ loudspeaker signal at frame index $\tau$ as $\bm{y}_l[\tau]$.

\subsection{Dictionary Generation}

We generate two dictionaries for a set of predefined positions, $\mathcal{S}$: First, a dictionary optimal fixed filters is computed for each listener position using the ACC/PM method, and stored as
\begin{align}
  \mathcal{D} = \left\{\bm{q}_l^{(s)}: s=1,\ldots,S,~l=1,\ldots,L \right\}, 
  \label{eq_control_filter_dictionary}
\end{align}
where $S=\lvert\mathcal{S}\rvert$ is the number of positions. Secondly, a dictionary of the IRs from the loudspeakers to the observation
microphones for each predefined position, i.e.,
\begin{flalign}
  &\mathcal{H}_O=\Bigl\{\bm{h}_{m_o,l}^{(s)}:
  l=1,\ldots,L,
  m_o=1,\ldots,M_O, s=1,\ldots,S \Bigr\}.\hspace{-4mm}&&
\end{flalign}
is stored and used for selection of the best filter from $\mathcal{D}$ when the listener position changes.

\subsection{Proposed Fixed Filter Selection with Audio-Only Tracking}
\label{subsection_similarity_matching}
For each input frame, $\bm{x}[\tau]$, we select the best SZC filter from the precomputed dictionary, $\mathcal{D}$, by comparing the sound frame recorded by the
observation microphones, ${\bm{p}_{m_o}[\tau] \in \mathbb{R}^{N}}$,
with internally generated sound frames at the observation
microphones for each listener position in the dictionary $\mathcal{H}_O$, at the current frame index $\tau$.
Using the loudspeaker signals, $\bm{y}_l[\tau]$,
the sound frame at observation microphone $m_o$ for position $s$ is estimated as
\begin{equation}
 \widetilde{\bm{z}}_{m_o}^{(s)}[\tau] = \sum\nolimits_{l=1}^{L}\bm{h}_{m_o,l}^{(s)} * \bm{y}_{l}[\tau] \in \mathbb{R}^{N+K-1}, \quad s \in \mathcal{S},
 \label{eq:estimated_sound_field}
\end{equation}
which has length $N+K-1$. In our formulation, we consider ${N+K-1 \leq 2N}$ and that all signal frames have length $N$.
The last $K-1$ samples from $\widetilde{\bm{z}}_{m_o}^{(s)}[\tau]$ are stored in the overlap-add buffer $\widetilde{\bm{z}}_{m_o, buff}^{(s)}[\tau]$ and the estimated sound frame at the observation microphone $m_o$ at frame index $\tau$, i.e., ${\bm{z}}_{m_o}^{(s)}[\tau]$, is obtained by adding the first $N$ samples of $\widetilde{\bm{z}}_{m_o}^{(s)}[\tau]$, to $\widetilde{\bm{z}}_{m_o,buff}^{(s)}[\tau-1]$.
%
%
We now compute the similarity between the observed and estimated sound frames at each observation microphone and for each dictionary position, using the NCS metric, $c^{(s)}_{m_o}[\tau]$, as\cite{tan_introduction_2014}
\begin{flalign}
    c^{(s)}_{m_o}[\tau] &= \left(\bm{p}_{m_o}[\tau]^{\textrm{T}}\bm{z}_{m_o}^{(s)}[\tau]\right)/\left(\lVert \bm{p}_{m_o}[\tau] \rVert_2 \lVert \bm{z}_{m_o}^{(s)}[\tau] \rVert_2\right);~s \in \mathcal{S},\hspace{-5mm}&&
    \label{eq:cosine_similarity}
\end{flalign}
which corresponds to the cosine of the angle between the two frames\cite{tan_introduction_2014}. We then sum the frame similarity over all observation
microphones to get the total signal similarity for each position, $s$, i.e.,
\begin{equation}
  c^{(s)}[\tau] = \sum\nolimits_{m_o=1}^{M_O} c^{(s)}_{m_o}[\tau].
\end{equation}
Finally, we select the listener position with the highest signal similarity as the estimate of listener position at current frame $\tau$, $s^*[\tau]$, i.e.,
\begin{equation}
  s^*[\tau] = \arg\max_{s} \left\{c^{(s)}[\tau]:s \in \mathcal{S}\right\}.
\end{equation}
The control filters for the next frame $\tau+1$ are then set to the control filters corresponding to the estimated listener position, $s^*[\tau]$, from the control filter dictionary $\mathcal{D}$, i.e.,
\begin{equation}
  \bm{q}_l[\tau+1] = \bm{q}_l^{\left(s^*[\tau]\right)},~~ \text{for}~l=1,\ldots,L, ~~ \text{where}~\bm{q}_l^{\left(s^*[\tau]\right)}\in \mathcal{D}.
\end{equation}
This procedure is repeated for each time frame $\tau$, to continuously monitor and adapt to changes in the listener position.

\section{Experiments}\label{sec:experiments}

\subsection{IR Measurement Setup}
\begin{figure}
    \centering
    \includegraphics[width=\linewidth]{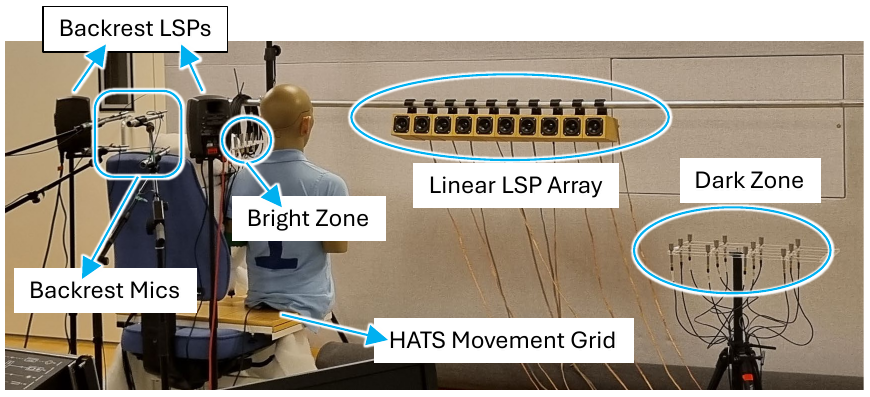}
    \caption{IR measurement setup.\vspace{-2mm}}
    \label{fig:measurement_setup}
\end{figure}

We evaluate performance using the SZC setup in Fig.~\ref{fig:measurement_setup} with a GRAS KEMAR $45$BC Head And Torso Simulator (HATS) with anthropometric pinnae\cite{gras_hats} to simulate a human listener. The setup is in a $60$\,m$^2$ room with ${T_{60} \approx 0.2}$\,s at Aalborg University (conforms to ITU-R BS775-1\cite{exp_room}). The BZ is sampled with $M_B=10$ microphones placed in a grid beside the left ear of the HATS, spaced $5$\,cm apart at a height of $129$\,cm. 
A large BZ beside one ear of the HATS is prioritized over two small BZs at either side to allow for a good listening experience if smaller undetectable movements are made within the BZ. 
The BZ is fixed to the HATS to represent changes in the true BZ as the listener moves. The HATS is set on a $60\,\textrm{cm}\times 40\,\textrm{cm}$ table with an office chair backrest. The table has $15$ grid points ($5$ along the $x$-axis and $3$ along the $y$-axis) spaced $10$\,cm apart, cf. Fig.~\ref{fig_grid_points}(a). At the top of the backrest, $2$ Genelec $8010$A loudspeakers are placed $46$\,cm apart, with $M_O=4$ microphones centered between them in a $2\times 2$ grid ($15$\,cm horizontal and $11$\,cm vertical spacing). A linear array of $10$, $2$\,inch loudspeakers in $10\,\textrm{cm}\times 10\,\textrm{cm}\times 16\,\textrm{cm}$ cabinets, spaced $10$\,cm apart, is set at a height of $121$\,cm and $170$\,cm in front of the table, aligned with its edge. Hence, the SZC system uses $L=10+2=12$ loudspeakers. The DZ is sampled with $M_D=12$ microphones, spaced $10$\,cm apart on a $2$D plane, centered $144$\,cm to the left of the table at a height of $102$\,cm. The HATS is moved across grid points, and at each point, all IRs from all the $L$ loudspeakers to the $M_B+M_D+M_O$ microphones are measured using the exponential sine sweep method \cite{farina2000simultaneous} at sampling frequency $f_s = 48$\,kHz, with a start frequency of $100$\,Hz and stop frequency of $f_s/2$. The IRs were truncated to a length of $K=4000$ samples to save computational complexity at inference time, since at this point their energy had decreased by more than $55$\,dB. 


\subsection{Reference Schemes}
We compare the proposed scheme with three reference schemes:\\
%
\noindent \textit{\textbf{(1) Mix Data Filter:} }
The BZs for all $S$ pre-measured listener positions are combined into into one large BZ and similarly for the DZ, as, 
${\bm{R}_B^{\textrm{mix}} = \frac{1}{S}\sum\nolimits_{s=1}^{S}\bm{R}_B^{(s)}}$ and ${\bm{R}_D^{\textrm{mix}} =\frac{1}{S} \sum\nolimits_{s=1}^{S}\bm{R}_D^{(s)}}$, respectively, and the optimal fixed filter, $\bm{q}^{\textrm{mix}}$, is derived using ACC/PM.
This is also equivalent to combining all pre-measured listener positions into a single large set of IRs 
and deriving the optimal fixed filter for this concatenated larger sound zone. 
The same fixed filter is then used for all listener positions. \\
%
\noindent\textit{\textbf{(2) Start Pos. Filter:} }
Traditional fixed filter SZC without any position tracking. Here we assume that the listener position is known only at the start and the filter stays fixed throughout the simulations. \\
%
\noindent\textit{\textbf{(3) Optimal filter:} }
Optimal fixed filter derived for the true listener position is used and provides the upper bound of performance.



\subsection{SZC With Position Tracking: Performance Comparison}

For all SZC methods and schemes, loudspeaker $l=6$ is the desired source, with control filters of length $J=1000$ samples and a frame size of $N=9600$ samples $=200$\,ms, and with $\zeta=0.5$ and $\lambda=1\times 10^{-5}$. These were chosen based on informal tuning of performance while considering memory and computational complexity. The input signal considered is a $48$\,kHz music signal with speech content \cite{bhattacharjee_low_complexity_2024}.
The performance metrics considered are: Time Domain Acoustic Contrast (TD AC), Time Domain normalized Signal Distortion (TD nSDP), Frequency Domain Acoustic Contrast (FD AC), and Frequency Domain normalized Signal Distortion (FD nSDP), as defined in \cite{lee2020fast, shi2021generation}.
The TD and FD nSDP for the ACC method is close to $0$ dB for all the schemes in both the evaluations below. Hence, for brevity we do not present nSDP for ACC here, but it can be found online\footnote[1]{\label{site_url}https://aerial-epoch-321.notion.site/HT-SZC-ac137859315e4a449fb2d3e8c76389ee?pvs=4}.

\begin{figure}
    \centering
    \includegraphics[width=\linewidth]{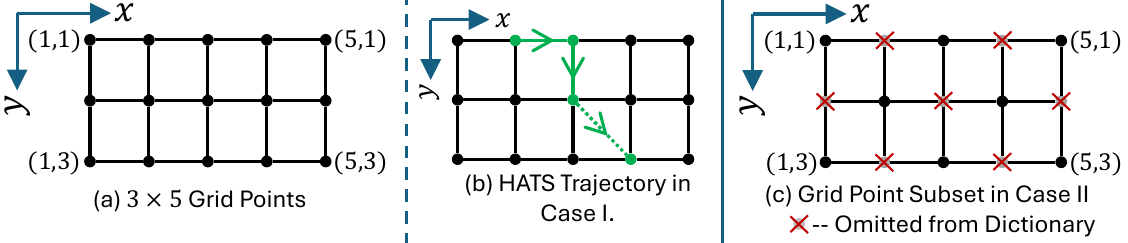}
    \caption{(a) IR measurement grid points $\mathcal{S}$; (b) HATS Trajectory in Case I; (c) $\bullet$ -- Grid points considered in $\mathcal{S}' \subset \mathcal{S}$ in Case II.
    \vspace{-2mm}}
    \label{fig_grid_points}
\end{figure}
\subsubsection{Case I: All Grid Points used in Dictionary}
To evaluate the merit of NCS as a similarity metric without any effect of unseen/unknown HATS positions, we consider the ideal case where the set of pre-measured positions, $\mathcal{S}$, is equal to the full grid of possible HATS positions shown in Fig.~\ref{fig_grid_points}(a). Thus, the dictionaries $\mathcal{D}$ and $\mathcal{H}_O$ contain filters and IRs for all grid positions, and the `Mix Data Filter', $\bm{q}^{\textrm{mix}}$, is derived from all grid positions. 
%
%
Fig. \ref{fig_grid_points}(b) shows the HATS trajectory, starting at position $(2,1)$ and changing positions three times at equal intervals during the simulation. 
The control filter for the proposed scheme is initialized as the `Mix Data Filter', i.e., $\bm{q}_l^{\textrm{mix}}$, and position tracking is performed at each frame index $\tau$, to select the control filter for frame $\tau+1$.

\begin{figure*}
\centering
\subfloat[Case I: Time domain performance. PC: HATS Position Change.  \label{fig_case_i_TD}]{\includegraphics[width=\linewidth]{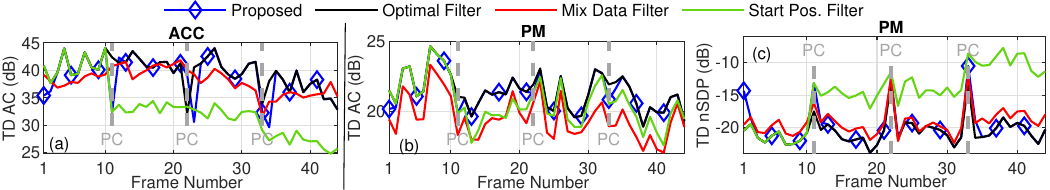}}\\
\subfloat[Case II: Time domain performance averaged over $50$ Monte-Carlo iterations. PC: HATS Position Change. \label{fig_case_2_TD}]{\includegraphics[trim={0 0 0 4mm}, clip, width=\linewidth]{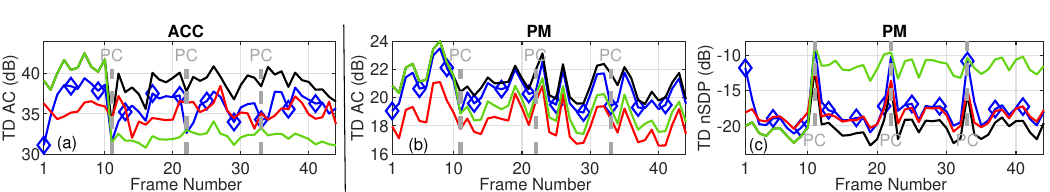}}\\
\subfloat[Case I: Frequency domain performance.\label{fig_case_i_FD}]{\includegraphics[trim={0 0 0 8.7mm}, clip,width=.48\linewidth]{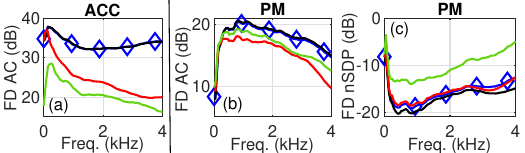}}
\hfill
\subfloat[Case II: Frequency domain performance averaged over $50$ Monte-Carlo iterations. \label{fig_case_2_FD}]{\includegraphics[trim={0 0 0 9mm}, clip, width=.48\linewidth]{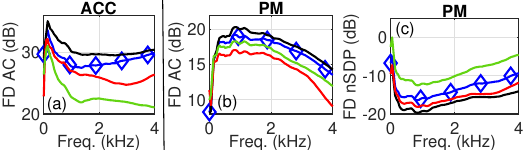}}
\caption{Comparison of time and frequency domain performance between the proposed and competing SZC schemes.\vspace{-2mm}}
\end{figure*}

The TD performance in Fig. \ref{fig_case_i_TD} shows, at the start position, the `Proposed' scheme quickly finds the correct grid position, matching the `Optimal Filter' performance for both ACC and PM methods. After the position changes, the `Start Pos. Filter' performance degrades significantly, highlighting the limitations of a single fixed filter. Conversely, the `Proposed' scheme estimates the correct HATS position within three frames after each change, maintaining performance on par with the `Optimal Filter'.
Throughout the simulation, the `Mix Data Filter' performs close to but slightly below the `Optimal Filter', yet much better than the `Start Pos. Filter' after position changes. Thus, the `Proposed' scheme performs the best. However, if no position tracking is done, the `Mix Data Filter' is a better choice than using the fixed start position filter to ensure good performance across different grid positions.
The FD performance in Fig. \ref{fig_case_i_FD}
is based on the reproduced signals in BZ and DZ over the entire simulation. Similar to the TD performance, the `Proposed' scheme closely matches the `Optimal Filter' performance, while the `Start Pos. Filter' performs the worst, and the `Mix Data Filter' scheme lies in between.
\renewcommand{\thesubfigure}{\roman{subfigure}}

\subsubsection{Case II: Subset of Grid Points used in Dictionary} 
We evaluate performance in a practical scenario where listener positions can lie outside the grid of pre-measured positions. Thus, the dictionaries $\mathcal{D}$ and $\mathcal{H}_O$ will not contain optimal filters and IRs for all possible positions, because the proposed method only uses a subset of grid points, $\mathcal{S}' \subset \mathcal{S}$, shown in Fig. \ref{fig_grid_points}(c), with black dots for included positions and red crosses for omitted ones. The `Mix Data Filter' $\bm{q}^{\textrm{mix}}$ is computed using only IRs from $\mathcal{S}'$. Again the initial filter for the proposed scheme is set equal to the `Mix Data Filter'.
Simulations run for $50$ Monte-Carlo iterations, with HATS positions changing three times per iteration, randomly picked from the $15$ grid points in Fig. \ref{fig_grid_points}(a). Thus, the start position may lie outside $\mathcal{S}'$, but the `Start Pos. Filter' always uses the optimal filter for the start position. TD and FD results are averaged over the $50$ iterations.

The TD results in Fig. \ref{fig_case_2_TD} show, that the `Start Pos. Filter' matches the `Optimal Filter' at the start but degrades significantly after position changes. The `Mix Data Filter' shows moderate improvement compared to the `Start Pos. Filter', especially in TD AC for ACC and TD nSDP for PM. Compared to the `Mix Data Filter', the `Proposed' scheme achieves a higher AC for both the AC and PM methods and similar nSDP.
The FD results in Fig. \ref{fig_case_2_FD} show, that the `Proposed' scheme achieves a higher AC than the `Start Pos. Filter' and `Mix Data Filter', and AC close to the `Optimal Filter'. In terms of nSDP, the `Proposed' and `Mix Data Filter' are close to the `Optimal Filter', while the `Start Pos. Filter' performs the worst. Both TD and FD results show that, even without optimal filters for all HATS positions, the proposed scheme provides better performance than using a single fixed filter by selecting a control filter solution from the nearby grid position available in $\mathcal{S}'$.

\section{Conclusion}\label{sec:conclusion}
This paper proposes a sound zone control (SZC) method that is robust to listener position changes, using a dictionary-based audio-only position tracking scheme. The method tracks the listener’s position via audio signals from observation microphones in the room, different from the bright and dark zone control points, and estimates the listener position using a dictionary of pre-measured impulse responses (IRs) for different positions, and then selects the optimal control filters for the estimated position.
Simulations with measured IRs show that when the dictionary includes IRs for all possible listener positions, the method accurately tracks the listener’s position and switches to the optimal filters, achieving the best control performance. Even when the listener moves to positions not available in the dictionary, the method improves SZC performance compared to using a single fixed filter at all times. 
This makes the scheme attractive for enhancing SZC robustness with a few observation microphones outside the controlled zones.
Finally, the presented method can be extended to include multiple environmental changes by pre-computing a dictionary covering a wider range of changes and then finding the correct IR among these in the same way.
Future studies will explore the effects of position grid density, noise at the observation
microphones, number of observation
microphones, and computational complexity.

\bibliographystyle{IEEEtran}
\bibliography{IEEEabrv_v_1_14, refs}

\begin{thebibliography}{10}
\providecommand{\url}[1]{#1}
\csname url@samestyle\endcsname
\providecommand{\newblock}{\relax}
\providecommand{\bibinfo}[2]{#2}
\providecommand{\BIBentrySTDinterwordspacing}{\spaceskip=0pt\relax}
\providecommand{\BIBentryALTinterwordstretchfactor}{4}
\providecommand{\BIBentryALTinterwordspacing}{\spaceskip=\fontdimen2\font plus
\BIBentryALTinterwordstretchfactor\fontdimen3\font minus \fontdimen4\font\relax}
\providecommand{\BIBforeignlanguage}[2]{{%
\expandafter\ifx\csname l@#1\endcsname\relax
\typeout{** WARNING: IEEEtran.bst: No hyphenation pattern has been}%
\typeout{** loaded for the language `#1'. Using the pattern for}%
\typeout{** the default language instead.}%
\else
\language=\csname l@#1\endcsname
\fi
#2}}
\providecommand{\BIBdecl}{\relax}
\BIBdecl

\bibitem{betlehem2015personal}
T.~Betlehem, W.~Zhang, M.~A. Poletti, and T.~D. Abhayapala, ``Personal sound zones: Delivering interface-free audio to multiple listeners,'' \emph{{IEEE} Signal Process. Mag.}, vol.~32, no.~2, pp. 81--91, 2015.

\bibitem{jeon2015active}
S.-W. Jeon, D.~H. Youn, Y.-c. Park, and G.-W. Lee, ``Active control of excessive sound emission on a mobile device,'' \emph{J. Acoust. Soc. Amer.}, vol. 137, no.~4, pp. EL327--EL333, 2015.

\bibitem{heuchel2020large}
F.~M. Heuchel, D.~Caviedes-Nozal, J.~Brunskog, F.~T. Agerkvist, and E.~Fernandez-Grande, ``Large-scale outdoor sound field control,'' \emph{J. Acoust. Soc. Amer.}, vol. 148, no.~4, pp. 2392--2402, 2020.

\bibitem{vindrola2021use}
L.~Vindrola, M.~Melon, J.-C. Chamard, and B.~Gazengel, ``Use of the filtered-x least-mean-squares algorithm to adapt personal sound zones in a car cabin,'' \emph{J. Acoust. Soc. Amer.}, vol. 150, no.~3, pp. 1779--1793, 2021.

\bibitem{galvez2014personal}
M.~F.~S. G{\'a}lvez, S.~J. Elliott, and J.~Cheer, ``Personal audio loudspeaker array as a complementary {TV} sound system for the hard of hearing,'' \emph{IEICE Trans. Fundam.}, vol.~97, no.~9, pp. 1824--1831, 2014.

\bibitem{jones2008personal}
M.~Jones and S.~J. Elliott, ``Personal audio with multiple dark zones,'' \emph{J. Acoust. Soc. Amer.}, vol. 124, no.~6, pp. 3497--3506, 2008.

\bibitem{cai2013design}
Y.~Cai, M.~Wu, and J.~Yang, ``Design of a time-domain acoustic contrast control for broadband input signals in personal audio systems,'' in \emph{Proc. {IEEE} Int. Conf. Acoust., Speech Signal Process.}\hskip 1em plus 0.5em minus 0.4em\relax IEEE, 2013, pp. 341--345.

\bibitem{poletti2008investigation}
M.~Poletti, ``An investigation of 2-{D} multizone surround sound systems,'' in \emph{Proc. 125th Audio Eng. Soc. Int. Conv.}, 2008.

\bibitem{olivieri2017generation}
F.~Olivieri, F.~M. Fazi, S.~Fontana, D.~Menzies, and P.~A. Nelson, ``Generation of private sound with a circular loudspeaker array and the weighted pressure matching method,'' \emph{{IEEE/ACM} Trans. Audio, Speech Lang. Process.}, vol.~25, no.~8, pp. 1579--1591, 2017.

\bibitem{chang_sound_2012}
J.-H. Chang and F.~Jacobsen, ``Sound field control with a circular double-layer array of loudspeakers,'' \emph{J. Acoust. Soc. Amer.}, vol. 131, no.~6, pp. 4518--4525, Jun. 2012.

\bibitem{lee2018unified}
T.~Lee, J.~K. Nielsen, J.~R. Jensen, and M.~G. Christensen, ``A unified approach to generating sound zones using variable span linear filters,'' in \emph{Proc. {IEEE} Int. Conf. Acoust., Speech Signal Process.}\hskip 1em plus 0.5em minus 0.4em\relax IEEE, 2018, pp. 491--495.

\bibitem{nielsen2018sound}
J.~K. Nielsen, T.~Lee, J.~R. Jensen, and M.~G. Christensen, ``Sound zones as an optimal filtering problem,'' in \emph{Proc. 52th Asilomar Conf. Signals, Syst. Comput.}, 2018, pp. 1075--1079.

\bibitem{shi2021generation}
L.~Shi, T.~Lee, L.~Zhang, J.~K. Nielsen, and M.~G. Christensen, ``Generation of personal sound zones with physical meaningful constraints and conjugate gradient method,'' \emph{{IEEE/ACM} Trans. Audio, Speech Lang. Process.}, vol.~29, pp. 823--837, 2021.

\bibitem{brunnstrom2022variable}
J.~Brunnstr{\"o}m, S.~Koyama, and M.~Moonen, ``Variable span trade-off filter for sound zone control with kernel interpolation weighting,'' in \emph{Proc. {IEEE} Int. Conf. Acoust., Speech Signal Process.}\hskip 1em plus 0.5em minus 0.4em\relax IEEE, 2022, pp. 1071--1075.

\bibitem{olsen_sound_2017}
M.~Olsen and M.~B. Møller, ``\BIBforeignlanguage{en}{Sound zones: on the effect of ambient temperature variations in feed-forward systems},'' in \emph{\BIBforeignlanguage{en}{Proc. 142nd Audio Eng. Soc. Int. Conv.}}\hskip 1em plus 0.5em minus 0.4em\relax Audio Engineering Society, 2017, pp. 1009--1018.

\bibitem{coleman_acoustic_2014}
P.~Coleman, P.~J.~B. Jackson, M.~Olik, M.~Møller, M.~Olsen, and J.~A. Pedersen, ``Acoustic contrast, planarity and robustness of sound zone methods using a circular loudspeaker array,'' \emph{J. Acoust. Soc. Amer.}, vol. 135, no.~4, pp. 1929--1940, Apr. 2014.

\bibitem{park_acoustic_2013}
J.-Y. Park, J.-W. Choi, and Y.-H. Kim, ``Acoustic contrast sensitivity to transfer function errors in the design of a personal audio system,'' \emph{J. Acoust. Soc. Amer.}, vol. 134, no.~1, pp. EL112--EL118, Jun. 2013.

\bibitem{jacobsen_living_2023}
R.~M. Jacobsen, K.~Fangel~Skov, S.~S. Johansen, M.~B. Skov, and J.~Kjeldskov, ``Living with {Sound} {Zones}: {A} {Long}-term {Field} {Study} of {Dynamic} {Sound} {Zones} in a {Domestic} {Context},'' in \emph{Proc. 2023 {CHI} {Conference} on {Human} {Factors} in {Computing} {Systems}}, Apr. 2023, pp. 1--14.

\bibitem{moller_moving_2020}
M.~B. Møller and J.~Østergaard, ``A {Moving} {Horizon} {Framework} for {Sound} {Zones},'' \emph{{IEEE/ACM} Trans. Audio, Speech Lang. Process.}, vol.~28, pp. 256--265, 2020.

\bibitem{elliott_robustness_2012}
S.~J. Elliott, J.~Cheer, J.-W. Choi, and Y.~Kim, ``Robustness and {Regularization} of {Personal} {Audio} {Systems},'' \emph{IEEE Trans. Audio, Speech, Lang. Process.}, vol.~20, no.~7, pp. 2123--2133, Sep. 2012.

\bibitem{han_combination_2019}
R.~Han, M.~Wu, C.~Gong, S.~Jia, T.~Han, H.~Sun, and J.~Yang, ``\BIBforeignlanguage{en}{Combination of {Robust} {Algorithm} and {Head}-{Tracking} for a {Feedforward} {Active} {Headrest}},'' \emph{\BIBforeignlanguage{en}{Appl. Sci.}}, vol.~9, no.~9, p. 1760, Apr. 2019.

\bibitem{jung_combining_2017}
W.~Jung, S.~J. Elliott, and J.~Cheer, ``\BIBforeignlanguage{en}{Combining the remote microphone technique with head-tracking for local active sound control},'' \emph{\BIBforeignlanguage{en}{J. Acoust. Soc. Amer.}}, vol. 142, no.~1, pp. 298--307, Jul. 2017.

\bibitem{elliott_head_2018}
S.~J. Elliott, W.~Jung, and J.~Cheer, ``\BIBforeignlanguage{en}{Head tracking extends local active control of broadband sound to higher frequencies},'' \emph{\BIBforeignlanguage{en}{Sci. Rep.}}, vol.~8, no.~1, p. 5403, Mar. 2018.

\bibitem{chang_multi-functional_2022}
C.-Y. Chang, C.-T. Chuang, S.~M. Kuo, and C.-H. Lin, ``\BIBforeignlanguage{en}{Multi-functional active noise control system on headrest of airplane seat},'' \emph{\BIBforeignlanguage{en}{Mech. Syst. Signal Process.}}, vol. 167, p. 108552, Mar. 2022.

\bibitem{xiao_ultra-broadband_2020}
T.~Xiao, X.~Qiu, and B.~Halkon, ``\BIBforeignlanguage{en}{Ultra-broadband local active noise control with remote acoustic sensing},'' \emph{\BIBforeignlanguage{en}{Sci. Rep.}}, vol.~10, no.~1, p. 20784, Nov. 2020.

\bibitem{buck_performance_2018}
J.~Buck, S.~Jukkert, and D.~Sachau, ``\BIBforeignlanguage{en}{Performance evaluation of an active headrest considering non-stationary broadband disturbances and head movement},'' \emph{\BIBforeignlanguage{en}{J. Acoust. Soc. Amer.}}, vol. 143, no.~5, pp. 2571--2579, May 2018.

\bibitem{shi_selective_2022}
D.~Shi, B.~Lam, K.~Ooi, X.~Shen, and W.-S. Gan, ``Selective fixed-filter active noise control based on convolutional neural network,'' \emph{Signal Process.}, vol. 190, p. 108317, Jan. 2022.

\bibitem{luo2018cosine}
C.~Luo, J.~Zhan, X.~Xue, L.~Wang, R.~Ren, and Q.~Yang, ``Cosine normalization: Using cosine similarity instead of dot product in neural networks,'' in \emph{Proc. Int. Conf. Artif. Neural Netw.}, 2018, pp. 382--391.

\bibitem{tan_introduction_2014}
P.-N. Tan, M.~Steinbach, and V.~Kumar, \emph{\BIBforeignlanguage{eng}{Introduction to Data Mining}}.\hskip 1em plus 0.5em minus 0.4em\relax Harlow: Pearson, 2014.

\bibitem{galvez2015time}
M.~F.~S. G{\'a}lvez, S.~J. Elliott, and J.~Cheer, ``Time domain optimization of filters used in a loudspeaker array for personal audio,'' \emph{{IEEE/ACM} Trans. Audio, Speech Lang. Process.}, vol.~23, no.~11, pp. 1869--1878, 2015.

\bibitem{moles2020personal}
V.~Mol{\'e}s-Cases, G.~Pi{\~n}ero, M.~de~Diego, and A.~Gonzalez, ``Personal sound zones by subband filtering and time domain optimization,'' \emph{{IEEE/ACM} Trans. Audio, Speech Lang. Process.}, vol.~28, pp. 2684--2696, 2020.

\bibitem{moller2016sound}
M.~B. M{\o}ller and M.~Olsen, ``Sound zones: On performance prediction of contrast control methods,'' in \emph{AES Int. Conf. Sound Field Control}, 2016.

\bibitem{bhattacharjee2023study}
S.~S. Bhattacharjee, L.~Shi, G.~Ping, X.~Shen, and M.~G. Christensen, ``Study and design of robust personal sound zones with {VAST} using low rank {RIRs},'' in \emph{Proc. {IEEE} Int. Conf. Acoust., Speech Signal Process.}\hskip 1em plus 0.5em minus 0.4em\relax IEEE, 2023, pp. 1--5.

\bibitem{christensen_introduction_2019}
M.~G. Christensen, \emph{\BIBforeignlanguage{en}{Introduction to {Audio} {Processing}}}.\hskip 1em plus 0.5em minus 0.4em\relax Cham: Springer International Publishing, 2019.

\bibitem{gras_hats}
\url{https://www.grasacoustics.com/products/head-torso-simulators-kemar/kemar-non-configured/product/749-45bc}, [Online; accessed 05-Aug-2024].

\bibitem{exp_room}
C.~V. Skipper, ``Listening room – multichannel,'' \url{https://vbn.aau.dk/en/equipments/listening-room-multichannel}, [Online; accessed 05-Aug-2024].

\bibitem{farina2000simultaneous}
A.~Farina, ``Simultaneous measurement of impulse response and distortion with a swept-sine technique,'' in \emph{Proc. 108th Audio Eng. Soc. Int. Conv.}, 2000.

\bibitem{bhattacharjee_low_complexity_2024}
S.~S. Bhattacharjee, A.~J. Fuglsig, J.~R. Jensen, L.~Shi, G.~Ping, H.~Shen, and M.~G. Christensen, ``\BIBforeignlanguage{English}{Low complexity signal adaptive sound zone control using subspace tracking},'' in \emph{\BIBforeignlanguage{English}{2024 International Workshop on Acoustic Signal Enhancement (IWAENC)}}, 2024.

\bibitem{lee2020fast}
T.~Lee, L.~Shi, J.~K. Nielsen, and M.~G. Christensen, ``Fast generation of sound zones using variable span trade-off filters in the {DFT}-domain,'' \emph{{IEEE/ACM} Trans. Audio, Speech Lang. Process.}, vol.~29, pp. 363--378, 2020.

\end{thebibliography}

\end{document}